\documentclass[aps,pra,twocolumn,superscriptaddress,a4paper,longbibliography]{revtex4-2}
\usepackage{graphicx}
\usepackage{natbib}
\usepackage{hyperref}
\hypersetup{
           breaklinks=true  
        }
\usepackage{amsmath}
\usepackage{mathtools}
\usepackage{braket}

\begin{document}


\title{Observation of the Mollow Triplet from an optically confined single atom}
\author{Boon Long Ng}
\affiliation{Center for Quantum Technologies, 3 Science Drive 2, Singapore 117543}
\author{Chang Hoong Chow}
\affiliation{Center for Quantum Technologies, 3 Science Drive 2, Singapore 117543}
\author{Christian Kurtsiefer}
\affiliation{Center for Quantum Technologies, 3 Science Drive 2, Singapore 117543}
\affiliation{Department of Physics, National University of Singapore, 2 Science Drive 3, Singapore 117542}
\date{\today}

\begin{abstract}
Resonance fluorescence from atomic systems consists of a single spectral peak that evolves into a Mollow triplet for a strong excitation field.
Photons from different peaks of the triplet show distinct photon correlation that make the fluorescence a useful light source for quantum information purpose.
We characterize the fluorescence of a single optically trapped $^{87}$Rb atom that is excited resonantly at different power levels.
Second-order correlation measurements reveal the single photon nature of the fluorescence concurrently with Rabi oscillations of a strongly excited atom.
The asymmetry in correlations between photons from two sidebands of the fluorescence spectrum when the atom is exposed to an off-resonant field further indicates that there is a preferred time-ordering of the emitted photons from different sidebands.
\end{abstract}


\maketitle
\section{Introduction}
The investigation of fluorescence emitted from resonantly excited atomic systems has played a major role in understanding the interaction between atom and radiation.
In 1930, Weisskopf first established the theory of atomic resonance fluorescence in the limit of weak excitation~\cite{Weisskopf1930}.
In this limit, the fluorescence spectrum of a two-level atom shows a single scattering peak centered at the excitation frequency.
Later this result was extended to include the effect of strong excitation radiation by Mollow in 1969~\cite{Mollow1969a}.
When the driving intensity increases above the saturation regime, the inelastic component in the fluorescence dominates, and the single peak spectrum evolves into a triplet structure.
The photons emitted in this process continue to be of interest in quantum optics, as these photons exhibit different correlation signatures in particular conditions such as off-resonant excitation~\cite{Cohen-TAnnoudji1979,Aspect1980a, Schrama1992, Nienhuis1993, Ulhaq2012a, Peiris2015, LopezCarreno2017}.

The Mollow triplet was first observed experimentally in an atomic beam passing perpendicularly through an intense laser field~\cite{Schuda1974, Hartig1976, Grove1977} where the emitted fluorescence spectrum was analyzed using a Fabry-Perot cavity.
This configuration minimized Doppler broadening due to atomic motion, and the fluorescence could be approximated as light emitted from individual non-interacting atoms.
Since then, the Mollow triplet has been successfully observed in many different systems such as quantum dots~\cite{Ulhaq2012a, Vamivakas2009a, Flagg2009a, Ates2009, Konthasinghe2012}, molecules~\cite{Wrigge2008a}, ions~\cite{Stalgies1996, Sterk2012}, cold atomic cloud~\cite{Ortiz-Gutierrez2019}, and superconducting qubits~\cite{Astafiev2010a, VanLoo2013, Toyli2016}.

While easier to implement experimentally, light interaction with an ensemble of atoms will mask certain features of the process such as photon anti-bunching.
In contrast, a single optically trapped atom is an excellent candidate to investigate photon correlations between different frequency components in the Mollow triplet.
An optically confined atom can be cooled to sub-Doppler temperature owing to polarization gradient cooling (PGC)~\cite{Ungar1989, Weiss1989}, and therefore suppresses the Doppler contribution to the spectrum.
Using a magnetic field to lift the Zeeman degeneracy and an appropriate driving laser polarization, the closed transition of an ideal two-level system can be implemented, coming close to the ideal situation considered in the Mollow triplet theory.

In this paper, we report the observation and analysis of fluorescence collected from a strongly driven single $^{87}$Rb atom in a far off-resonance optical dipole trap (FORT).
An aspherical lens focuses near-resonant probe laser light onto the atom and collects backscattered photons with minimal laser background.
The probe is near-resonant with the closed transition $5S_{1/2}\ket{F=2,\ m_{F}=-2} \equiv \ket{g}$ to $5P_{3/2}\ket{F=3,\ m_{F}=-3} \equiv \ket{e}$.
We analyze the spectrum of the light scattered by the atom at different excitation intensities with a scanning Fabry-Perot cavity.
A second-order photon correlation measurement of the fluorescence shows the signature Rabi oscillation with frequency that relates to the driving intensity.
Under off-resonant excitation, the temporal cross-correlation between photons originating from different sidebands is measured to reveal the dynamics of the underlying optical transitions.

\section{Theoretical background}\label{sec:mollow}
In the limit of a strong driving field $(\Omega \gg \Gamma /4)$, the power spectrum of the resonance fluorescence is given by
\begin{equation}
\label{eqn:mollow_spec}
\begin{split}
S(\omega) = \dfrac{\Gamma / 4\pi}{\omega^{2} + (\Gamma/2)^{2}}& + \dfrac{3\Gamma / 16\pi}{(\omega+\Omega)^{2}+(3\Gamma/4)^{2}} \\
& + \dfrac{3\Gamma / 16\pi}{(\omega-\Omega)^{2}+(3\Gamma/4)^{2}}\,,
\end{split}
\end{equation}
where $\omega$ is the relative frequency from the monochromatic driving field, $\Omega$ is the Rabi frequency and $\Gamma$ represents the natural linewidth of the atomic transition, which in this case is $2\pi \times 6.07$\,MHz for $^{87}$Rb D2 transition.

According to Eqn.~(\ref{eqn:mollow_spec}), the power spectrum contains a resonant Lorentzian peak with a full width half maximum (FWHM) of $\Gamma$ and two side peaks located at $\pm \Omega$ away from the resonance, with a FWHM of $3 \Gamma /2$.
These sidebands, together with the central peak, form the Mollow triplet.
This result was derived by Mollow using a semi-classical approach~\cite{Mollow1969a}, but the same result can be obtained using a fully quantum-mechanical picture~\cite{Kimble1976}.

One way to interpret the spectral features is to describe the atomic energy states as dressed by the driving field~\cite{Cohen-Tannoudji1977, Nienhuis1993}.
In this dressed-state picture, the new eigenstates are a superposition of the bare states $\ket{g, n+1}$ and $\ket{e, n}$, where ``g" and ``e" refer to the ground and excited states of the atom, while $n$ indicates the number of photons from the driving field.
In every manifold where the total number $N$ of photons plus atomic excitation is the same, the eigenstates are split by the Rabi frequency.

The three frequency components in the fluorescence can be explained by spontaneous decay from a manifold of $N$ total excitations to a manifold with $(N-1)$ excitations.
Four optical transitions are possible in this process.
Two of them are degenerate and correspond to the central peak in the fluorescence spectrum, while the sidebands $\pm \Omega$ away from the central peak originate from the other two transitions.
This leads to the weightage of 1:2:1 in the total spectral intensities of the incoherent peaks under resonant excitation.

\begin{figure}
\centering
  \includegraphics[width=\columnwidth]{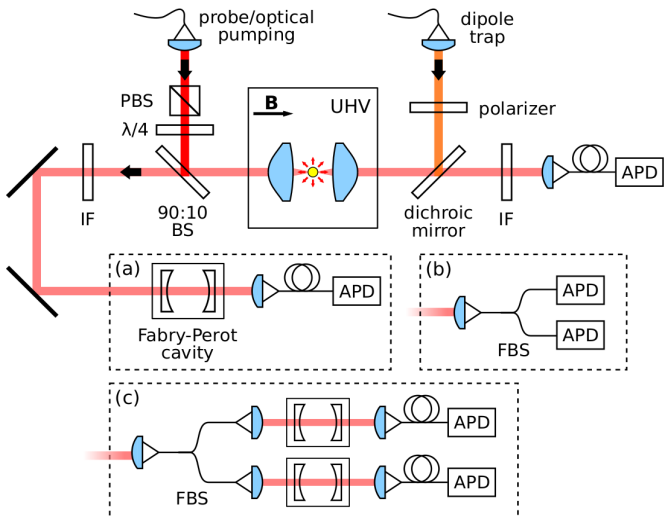}
  \caption{\label{fig:setup} Setup for probing light-atom interaction in free
    space. A single $^{87}$Rb atom is cooled and trapped in a far off-resonance
    dipole trap. One avalanche photodetector (APD) is used to monitor the atomic fluorescence which act
    as a trigger to start experiment sequence. (a) A Fabry-Perot cavity is
    placed before another APD to measure the frequency spectrum of the atomic
    fluorescence. (b) Hanbury Brown and Twiss (HBT) configuration to measure
    second-order intensity autocorrelation. (c) Cross-correlation measurement
    setup by placing a cavity in each arm before each APD to select photon from
    specific frequency window.
    UHV:~ultra-high vacuum chamber, 
  IF:~interference filter centered at 780\,nm, $\lambda$/4:~quarter-wave plate, PBS:~polarizing beam splitter, FBS:~fiber beam splitter, B:~magnetic field.
}
\end{figure}

\section{Experimental setup}\label{sec:setup}
Our experiment starts with a single $^{87}$Rb atom trapped in a red-detuned FORT that is loaded from a magneto-optical trap (MOT) (see Fig.~\ref{fig:setup}).
This dipole trap is formed by a linearly polarized Gaussian laser beam (wavelength 851\,nm) that is tightly focused by a pair of high numerical aperture lenses (NA = 0.75, focal length $f$ = 5.95\,mm) to a waist of $w_{0} = 1.1$\,$\mu$m.
Part of the atomic fluorescence is collected through the same lenses and coupled into single mode fibers that are connected to avalanche photodetectors (APD).

Once an atom is trapped, we apply 10\,ms of PGC to reduce the atomic motion to a temperature of 14.7(2)\,$\mu$K~\cite{chin2017PGC}.
Then, a bias magnetic field of 1.44\,mT is applied along the FORT laser propagation direction to remove the degeneracy of the Zeeman states, and the atom is optically pumped into $\ket{g}$~\cite{chow2021}.
Next, we turn on the probe laser beam along the optical axis for 2\,$\mu$s.
This length is chosen to maximize the duty cycle of photon collection while avoiding excessive recoil heating of the atom.
The probe frequency is locked to the $F=2 \rightarrow F'=3$ hyperfine transition of the $^{87}$Rb $D_{2}$ line, and shifted by an acousto-optic modulator (AOM) in order to address the $\ket{g} \leftrightarrow \ket{e}$ transition.
The probe is prepared into a $\sigma^{-}$ polarization with a quarter-wave plate after a polarizing beam splitter (PBS) to target the closed transition.

\begin{figure}
\centering
  \includegraphics[width=\columnwidth]{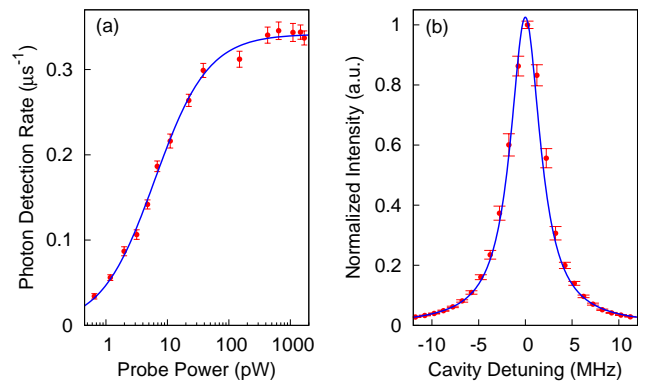}
  \caption{\label{fig:characterization} (a) Resonant saturation measurement,
    with the blue solid line representing the fit to $\dfrac{\eta \Gamma}{2}
    \dfrac{P_{\text{probe}}}{P_{\text{probe}}+P_{\text{sat}}}$ with saturation
    power $P_{\text{sat}}=$ 6.3(2)\,pW and total detection efficiency $\eta=$
    1.79(2)\,\%. Here, $P_{\text{probe}}$ is incident probe power. (b) Cavity transmission of the probe laser to characterize the cavity linewidth.
}
\end{figure}

We collect photons scattered backwards through the same lens and couple them into a single mode fiber, avoiding the strong light levels of the probe laser for analysis.
The photon scattering rate is first characterized for different intensity levels of the probe field, as illustrated in Fig.~\ref{fig:characterization}(a).
The atomic response saturates at a probe power of 6.3(2)\,pW and total detection efficiency, $\eta=$ 1.79(2)\,\% can be inferred from the fit.

The collected photons are frequency-filtered with a Fabry-Perot cavity
and subsequently detected with an APD.
By scanning the cavity resonance frequency, the frequency spectrum of the
fluorescence can be obtained.
To precisely control the resonance frequency of this cavity, it is locked to a tunable sideband generated by an electro-optical modulator (EOM) from another laser locked to the D1 transition of $^{87}$Rb.
The linewidth of the cavity is characterized to be 3.92(5)\,MHz with an external cavity diode laser (see Fig.~\ref{fig:characterization}(b)).
This value will be used for deconvolution of the atomic spectrum in the next part of this paper.

\begin{figure}
\centering
  \includegraphics[width=\columnwidth]{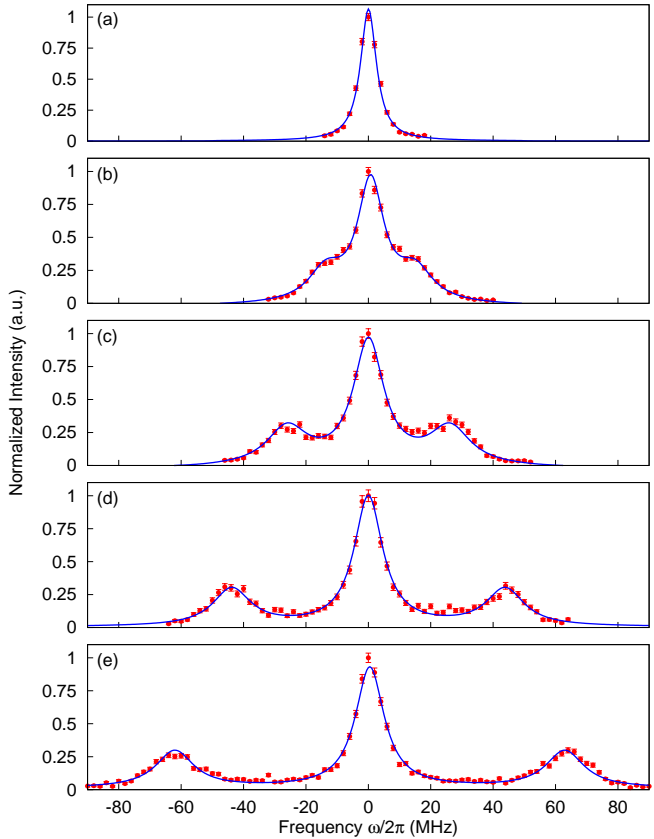}
  \caption{\label{fig:spectrum} Normalized resonance atomic emission spectra
    at different excitation intensities recorded by scanning the Fabry-Perot
    cavity with the setup in Fig.~\ref{fig:setup}(a). For (b)-(e), solid line is a fit to Eqn.~(\ref{eqn:mollow_spec}) convoluted with the cavity transfer function and the effect of laser reflection.
}
\end{figure}

\begin{figure}
\centering
  \includegraphics[width=\columnwidth]{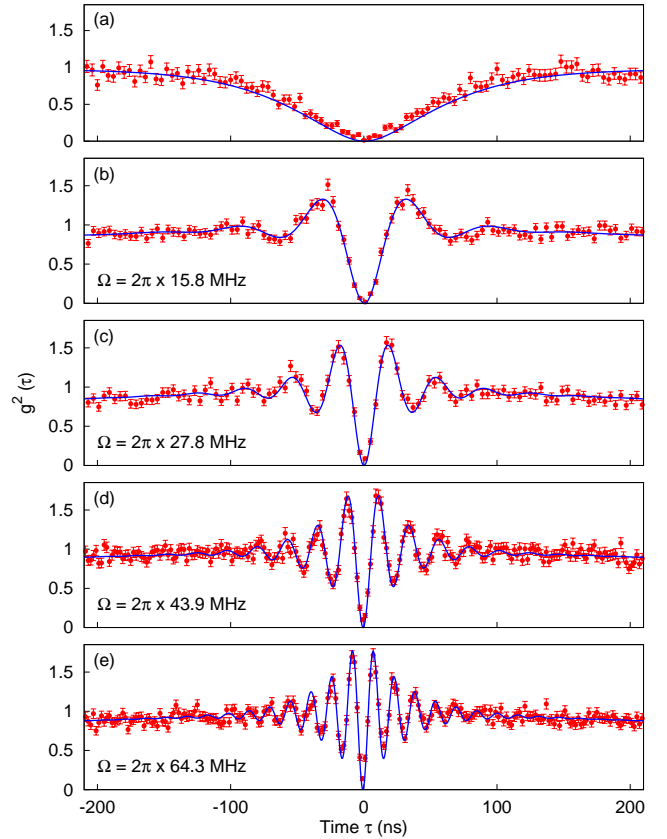}
  \caption{\label{fig:g2} Second-order correlation function of the single atom at different excitation intensities. Solid line is a fit to Eqn.~(\ref{eqn:g2_atom}) with addition of triangle function resulted from a convolution of two square pulses. The Rabi frequency $\Omega$ extracted from the fit is labeled in Fig.~\ref{fig:g2}(b)-(e).
}
\end{figure}

Figure~\ref{fig:spectrum} shows a series of frequency spectra for increasing excitation powers.
At weak excitation, the FWHM of the single peak in Fig.~\ref{fig:spectrum}(a) is 2.5(3)\,MHz after deconvolution from the cavity contribution.
This shows that at a driving power that is well below saturation, the coherent component with linewidth smaller than $\Gamma$ dominates the spectrum.

As the power increases, the three-peak structure emerges and the splitting between the peaks also increases.
The fit to the experimental data is done with Eqn.~(\ref{eqn:mollow_spec}) convoluted with the cavity transfer function.
After excluding cavity contribution, the central peak in Fig.~\ref{fig:spectrum}(e) has a FWHM of 7.3(5)\,MHz extracted from the fit.
This value is close to the atomic natural linewidth of $^{87}$Rb, thus justifying the claim that an optically trapped single atom can be laser cooled to mitigate the Doppler broadening effect.
Theoretically, the height ratio between the central peak and the sidebands is 1\,:\,3\,:\,1 according to Eqn.~(\ref{eqn:mollow_spec}).
After taking into account the cavity contribution, the height of the central
peak should decrease such that the ratio reaches around 1\,:\,2.6\,:\,1.
However, the measured spectra show central peaks with
about 3.7 times the height of sidebands (average value of Fig.~\ref{fig:spectrum}(c)-(e)).
This inconsistency between the theoretical prediction and the experimental
data can be likely attributed to the reflection and scattering of the probe laser from the optics.
We characterized this laser reflection and found a contribution of around
7.6\,\% of the total power in the spectrum.
With addition of this contribution, the calculation shows that the ratio after cavity convolution will be 1\,:\,3.7\,:\,1\,, which explain the measurement results.

\section{Second Order Correlation Function}\label{sec:g2}
In the subsequent part of the experiment, we replace the Fabry-Perot cavity with a fiber beam splitter and two APDs in a Hanbury-Brown and Twiss configuration as shown in Fig.~\ref{fig:setup}(b)~\cite{Brown1956}.
The arrival time of the photons is recorded.
The second-order intensity correlation function ($g^{(2)}(\tau)$) of the
atomic fluorescence can be inferred from this measurement.
This correlation function can reveal some characteristics of the photons emitted by single atom such as photon anti-bunching.
It was first demonstrated experimentally by Kimble $et$ $al.$ in 1977~\cite{Kimble1977a} which showed that fluorescence from a two-level atom is manifestly quantum.
While a vanishing second-order intensity correlation of the fluorescence at
zero delay is a clear indication for this phenomenon, the dynamic of
$g^{(2)}(\tau)$ near the zero delay reveals more about the underlying
atom-light interaction such as a Rabi oscillation.

For driving fields of low intensity, $g^{(2)}(\tau)$ shows a monotonic increase to unity as $\tau$ increases from zero to much larger than $1/\Gamma$.
When the driving field intensity increases above the saturation, $g^{(2)}(\tau)$ resembles the case for weak excitations at large delay, but oscillations corresponding to the Rabi frequency appear around zero delay.
Upon the detection of first fluorescence photon, the atom is being projected onto the ground state and the probability to detect the subsequent photon at some later time, $\tau$ is proportional to the excited state population of the atom.
This correlation function for fluorescence from a single atom can be expressed as~\cite{loudon2000quantum}
\begin{equation}
\label{eqn:g2_atom}
g^{(2)}(\tau) = 1 - e^{-(3\Gamma /4)|\tau|} \left( \cos \Omega \tau + \dfrac{3\Gamma}{4\Omega} \sin \Omega |\tau| \right)\,.
\end{equation}

The correlation measurements shown in Fig.~\ref{fig:g2} are fitted using
Eqn.~(\ref{eqn:g2_atom}) multiplied with a triangle function that results from
convolution of two square pulses of the same length.
The Rabi frequency $\Omega$ can be also extracted from the fit, and it serves
as an independent measurement allowing comparison to the values obtained from
the Mollow triplet measurement.
The extracted values, shown in Fig.~\ref{fig:g2} for different driving powers,
agree well with the values for $\Omega$ obtained from the Mollow triplet
spectra.

\begin{figure}
\centering
  \includegraphics[width=\columnwidth]{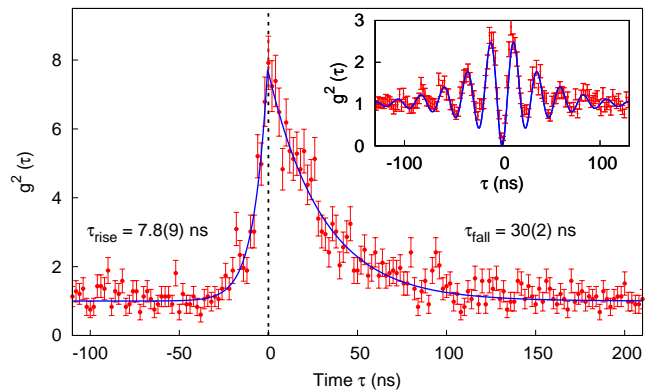}
  \caption{\label{fig:off_reso_g2} Normalized cross-correlation between
    photons from two opposite Mollow sidebands as a function of delay $\tau$
    between detection of a photon from the higher energy sideband after
    detection of a photon from lower energy sideband. Inset: Normalized intensity autocorrelation of the unfiltered off-resonance atomic fluorescence to extract $\Omega'$.
}
\end{figure}

\section{Off-resonant excitation}\label{sec:off}
While the atom is excited resonantly, the emission of the sideband photons does not have a preferred order.
As such, the cross-correlation between photons from different sidebands is symmetric with respect to zero time delay, $\tau = 0$.
However, if the excitation field is detuned from the atomic resonance, this symmetry is broken as the emission process of the sideband photons now have a preferred order~\cite{Aspect1980a, Schrama1992, Nienhuis1993, Ulhaq2012a}.
The preferred order of the emission depends on the sign of the detuning and
manifests as an asymmetry in the correlation measurement around $\tau = 0$.

In this part of the experiment, we red-detune the excitation laser by $30$\,MHz from the atomic resonance.
As shown in Fig.~\ref{fig:setup}(c), there is a Fabry-Perot cavity in front of each APD to filter the incoming fluorescence such that photon correlation between chosen spectral components can be measured.
To better transmit the photons from different peaks, the cavities used in this experiment have linewidth of $20$\,MHz.

The spectrum of the fluorescence is slightly different when the atom is
excited off-resonantly, with the central peak sitting at the driving
frequency and the sideband are separated from central peak by the generalized Rabi frequency, $\Omega' =  \sqrt{\Omega^{2} + \Delta^{2}}$ where $\Delta$ is the detuning of the laser from atomic resonance.
In order to align the cavity resonance with the respective sidebands, we first
measure the second-order correlation of the off-resonance fluorescence.
The data is shown in the inset of Fig.~\ref{fig:off_reso_g2} and the blue solid line is the fit to extract $\Omega'$, which is $2\pi\times 42(1)$\,MHz in this case.
As such, the cavity resonance is locked at $\pm\Omega'$ away from the driving frequency to isolate the sidebands photon.

Figure~\ref{fig:off_reso_g2} shows the cross-correlation measurement between the opposite Mollow sidebands where we use photon from the lower energy sideband as `start' trigger and the photon from the other sideband as `stop' signal.
The measurement shows a clear bunching behavior around $\tau = 0$.
The normalized correlation is fitted by two exponentials, with time constants of $\tau_{\text{rise}}=7.8(9)$\,ns and $\tau_{\text{fall}}=30(2)$\,ns, respectively.
The asymmetry of the correlation function justifies the claim that the emission of the sidebands photon has a preferred time-order for off-resonant excitation, which in this case photon emission from the lower energy sideband followed by photon emission from the higher energy sideband forms a temporal cascade emission.

\section{Conclusion}\label{sec:con}
In summary, we have measured the frequency spectrum of the resonance fluorescence of an optically trapped atom at different excitation intensities, until the emitter is saturated.
The distinctive Mollow triplet has been observed and has been compared to the theoretical model.
After taking into account the effect of the cavity transfer function and excitation power fluctuations, our results agree with the theoretical prediction very well.
For each excitation intensity used in the measurements of the emission spectra, we also record the second order correlation function of the atomic fluorescence.
The Rabi frequency can be extracted by fitting $g^{(2)}(\tau)$ and this value
serves as a benchmark for the results obtained in each measured spectrum.
With off-resonant excitation, the photons from opposite sidebands have a preferred order of emission which is reflected in the asymmetry of the correlation around $\tau = 0$.
This can be useful when using a single atom as a heralded narrowband single
photon source, or a quantum resource in quantum network where the quantum
information is stored and processed through a node, which could be atomic ensemble~\cite{Kuzmich2003} or a single atom within a cavity\cite{tatjana2007,Ritter2012}.

\begin{acknowledgments}
We acknowledge support of this work by the Ministry of Education in
Singapore and the National Research Foundation, Prime Minister's
office, through the Research Centres of Excellence programme.
\end{acknowledgments}

\bibliography{mollow.bib}{}

\end{document}